\documentclass[10pt,hidelinks,conference]{IEEEtran}

\ifCLASSINFOpdf
\else
\usepackage[dvips]{graphicx}
\fi
\usepackage{url}

\usepackage{graphicx} % Allows including images
\usepackage{booktabs} % Allows the use of \toprule, \midrule and \bottomrule in tables
\usepackage{mathrsfs}
\usepackage{amsmath}
\usepackage{amssymb}
\usepackage{bm}
\usepackage{mathtools}
\usepackage{cite}
\usepackage[acronym,shortcuts]{glossaries}
\usepackage{enumitem} % Customized lists
\usepackage{comment}
\usepackage{algorithm, algpseudocode}
\usepackage{mathtools}
\usepackage{lipsum}
\usepackage{mleftright}
\usepackage{orcidlink}
\usepackage{arydshln}
\usepackage{subcaption}

\setlist[itemize]{noitemsep} % Make itemize lists more compact

\DeclareMathOperator*{\argmin}{arg\,min}

\newcommand{\C}{{\bf C}}

\providecommand{\UL}{\mathrel{\raise-4pt\hbox{\hglue -2.8ex
\vrule height .1ex width 2.3ex
\vrule height 3ex width .1ex
\hglue .4ex}}}
\providecommand{\ul}{\mathrel{\raise-2pt\hbox{\hglue -2.3ex
\vrule height .1ex width 2ex
\vrule height 2ex width .1ex
\hglue .4ex}}}
\providecommand{\ub}{
\mathrel{\hbox{\hglue -2.8ex \vrule height 2ex width .06ex}
\raise2ex\hbox{\hglue -0.1ex\vrule height .1ex width 2.5ex}
\hbox{\hglue -0.1ex \vrule height 2ex width .1ex
\hglue .4ex}}}
\providecommand{\lb}{
\mathrel{\raise-0.5ex
\hbox{\hglue -2.8ex \vrule height 2ex width .1ex
\vrule height .1ex width 2.5ex
\vrule height 2ex width .1ex
\hglue .4ex}}}
\providecommand{\UB}{
\mathrel{
\raise-1ex\hbox{\hglue -1.8em \vrule height 3ex width .1ex}
\raise2ex\hbox{\hglue -0.1ex\vrule height .1ex width 1.5em}
\raise-1ex\hbox{\hglue -0.1ex \vrule height 3ex width .1ex
\hglue .4ex}}}
\providecommand{\LB}{
\mathrel{\raise-1ex
\hbox{\hglue -1.8em \vrule height 3ex width .1ex
\vrule height .1ex width 1.5em
\vrule height 3ex width .1ex
\hglue .4ex}}}
\usepackage{trimclip}
\newif\iflclip
\newif\ifbclip
\newif\ifrclip
\newif\iftclip
\def\CLIP{\dimexpr\fboxrule+.2pt\relax}
\def\nulclip{0pt}
\newcommand\partbox[2]{%
\lclipfalse\bclipfalse\rclipfalse\tclipfalse%
\let\lkern\relax\let\rkern\relax%
\let\lclip\nulclip\let\bclip\nulclip\let\rclip\nulclip\let\tclip\nulclip%
\parseclip#1\relax\relax%
\iflclip\def\lkern{\kern\CLIP}\def\lclip{\CLIP}\fi
\ifbclip\def\bclip{\CLIP}\fi
\ifrclip\def\rkern{\kern\CLIP}\def\rclip{\CLIP}\fi
\iftclip\def\tclip{\CLIP}\fi
\lkern\clipbox{\lclip{} \bclip{} \rclip{} \tclip}{\fbox{#2}}\rkern%
}
\def\parseclip#1#2\relax{%
\ifx l#1\lcliptrue\else
\ifx b#1\bcliptrue\else
\ifx r#1\rcliptrue\else
\ifx t#1\tcliptrue\else
\fi\fi\fi\fi
\ifx\relax#2\relax\else\parseclip#2\relax\fi
}
\usepackage{efbox}

\newacronym{SVD}{SVD}{singular value decomposition}
\newacronym{DCM}{DCM}{double-centering matrix}
\newacronym{3D}{3D}{three-dimensional}
\newacronym{GA}{GA}{genie-aided}
\newacronym{EA}{EA}{``\emph{estimate-then-average}''}
\newacronym{AE}{AE}{``\emph{average-then-estimate}''}
\newacronym{IRS}{IRS}{intelligent reflecting surface}
\newacronym{RSSI}{RSSI}{received signal strength indicator}
\newacronym{SotA}{SotA}{state-of-the-art}
\newacronym{CSI}{CSI}{channel state information}
\newacronym{D2D}{D2D}{device-to-device}
\newacronym{RR}{RR}{round-robin}
\newacronym{DA}{DA}{Dutch auction}
\newacronym{AV}{AV}{autonomous vehicle}
\newacronym{CWFL}{CWFL}{clustered WFL}
\newacronym{WFL}{WFL}{wireless federated learning}
\newacronym{RSMA}{RSMA}{rate splitting multiple access}
\newacronym{IoT}{IoT}{Internet-of-Things}
\newacronym{TDMA}{TDMA}{time-domain multiple access}
\newacronym{NOMA}{NOMA}{non-orthogonal multiple access}
\newacronym{ML}{ML}{machine learning}
\newacronym{MIMO}{MIMO}{multiple-input multiple-output}
\newacronym{CT}{CT}{compute-then-transmit}
\newacronym{FP}{FP}{fractional programming}
\newacronym{CF-mMIMO}{CF-mMIMO}{cell free massive MIMO}
\newacronym{iid}{i.i.d.}{independent and identically distributed}
\newacronym{AD}{AD}{autonomous driving}
\newacronym{DL}{DL}{downlink}
\newacronym{UL}{UL}{uplink}
\newacronym{IC}{IC}{interference cancellation}
\newacronym{SIC}{SIC}{successive interference cancellation}
\newacronym{BS}{BS}{base station}
\newacronym{TX}{TX}{transmit}
\newacronym{RX}{RX}{receive}
\newacronym{MU}{MU}{multi-user}
\newacronym{SISO}{SISO}{single-input single-output}
\newacronym{AWGN}{AWGN}{additive white Gaussian noise}
\newacronym{SINR}{SINR}{signal-to-interference-and-noise ratio}
\newacronym{FL}{FL}{federated learning}
\newacronym{CPU}{CPU}{central processing unit}
\newacronym{KNN}{KNN}{K-nearest-neighbor}
\newacronym{RF}{RF}{radio frequency}
\newacronym{GD}{GD}{gradient descent}
\newacronym{V2X}{V2X}{vehicle-to-anything}

\newacronym{RSS}{RSS}{received signal strength}
\newacronym{FIM}{FIM}{fisher information matrix}
\newacronym{ToA}{ToA}{time of arrival}
\newacronym{ToF}{ToF}{time of flightl}
\newacronym{AoA}{AoA}{angle of arrival}
\newacronym{GP}{GP}{Gaussian process}
\newacronym{2D}{2D}{two-dimensional}
\newacronym{GPR}{GPR}{Gaussian process regression}
\newacronym{GNSS}{GNSS}{global navigation satellite systems}
\newacronym{B5G}{B5G}{beyond fifth-generation}
\newacronym{6G}{6G}{sixth-generation}

\newacronym{RRH}{RRH}{remote radio head}
\newacronym{GPS}{GPS}{Global Positioning System}
\newacronym{RFID}{RFID}{radio frequency identification}
\newacronym{TCAS}{TCAS}{traffic alert and collision avoidance systems}
\newacronym{RMSE}{RMSE}{root mean square error}
\newacronym{MSE}{MSE}{mean square error}

\newacronym{SGD}{SGD}{stochastic gradient descent}
\newacronym{PDF}{PDF}{probability density function}
\newacronym{CU}{CU}{computing unit}
\newacronym{DM-MIMO}{DM-MIMO}{distributed massive multiple-input multiple-output}
\newacronym{LOS}{LOS}{line-of-sight}
\newacronym{NLOS}{NLOS}{non-line-of-sight}
\newacronym{ROI}{ROI}{region of interest}
\newacronym{AP}{AP}{access point}
\newacronym{TDOA}{TDOA}{time difference of arrival}
\newacronym{UE}{UE}{user equipment}
\newacronym{dB}{dB}{decibel}
\newacronym{RIS}{RIS}{reconfigurable intelligent surface}
\newacronym{CG}{CG}{conjugate gradient}

\newacronym{PG}{PG}{proximal gradient}
\newacronym{SVT}{SVT}{singular value thresholding}
\newacronym{NN}{NN}{nuclear norm}
\newacronym{NMSE}{NMSE}{normalized mean square error}
\newacronym{MC}{MC}{matrix completion}
\newacronym{NP}{NP}{non-deterministic polynomial-time}
\newacronym{EDM}{EDM}{euclidean distance matrix}
\newacronym{SC}{SC}{soft-connected}
\newacronym{CRLB}{CRLB}{Cramér-Rao Lower Bound}
\newacronym{PoA}{PoA}{phase of arrival}
\newacronym{UAV}{UAV}{unmanned aerial vehicle}
\newacronym{VR}{VR}{virtual reality}
\newacronym{MDS}{MDS}{multidimensional scaling}
\newacronym{SMDS}{SMDS}{super multidimensional scaling}

\newacronym{RBL}{RBL}{rigid body localization}
\newacronym{RBT}{RBT}{rigid body tracking}
\newacronym{SC-RBL}{SC-RBL}{soft-connected RBL}
\newacronym{W-RBL}{W-RBL}{\underline{wireless} RBL}

\newacronym{SDP}{SDP}{semidefinite programming}
\newacronym{JCAS}{JCAS}{joint communication and sensing}
\newacronym{SDR}{SDR}{semi-definite relaxation}

\newacronym{OPP}{OPP}{orthogonal Procrustes problem}
\newacronym{SLAM}{SLAM}{simultaneous localization and mapping}
\newacronym{WLS}{WLS}{weighted least square}
\newacronym{SI}{SI}{soft-impute}
\newacronym{GaBP}{GaBP}{Gaussian belief propagation}

\begin{document}

\title{SMDS-based Rigid Body Localization\vspace{-.5ex}}

\author{\IEEEauthorblockN{Niclas~F\"uhrling$^\dag$\textsuperscript{\orcidlink{0000-0003-1942-8691}}, Giuseppe Abreu$^\dag$\textsuperscript{\orcidlink{0000-0002-5018-8174}}, David~Gonz{\'a}lez~G.$^\ddag$\textsuperscript{\orcidlink{0000-0003-2090-8481}} and Osvaldo~Gonsa$^\ddag$\textsuperscript{\orcidlink{0000-0001-5452-8159}}}
\IEEEauthorblockA{\textit{$^\dag$School of Computer Science and Engineering, Constructor University, Bremen, Germany} \\ \textit{$^\ddag$Wireless Communications Technologies Group, Continental AG, Frankfurt, Germany} \\ 
(nfuehrling, gabreu)@constructor.university, david.gonzalez.g@ieee.org, osvaldo.gonsa@continental-corporation.com\\[-1ex]}
}

\setlength{\parskip}{0pt}

\maketitle

\begin{abstract}
We consider a novel \ac{RBL} method, based only on a set of measurements of the distances, as well as the angles between sensors of the vehicle to the anchor landmark points.
A key point of the proposed method is to use a variation of the \ac{SMDS} algorithm, where only a minor part of the complex edge kernel is used, based on the available information, which in the case of \ac{RBL} is anchor-to-anchor and target-to-target information.
Simulation results illustrate the good performance of the proposed technique in terms of \ac{MSE} of the estimates, compared also to the corresponding \ac{CRLB}.
\end{abstract}

\IEEEpeerreviewmaketitle

\vspace{-3ex}
\section{Introduction}
\vspace{-1ex}

\IEEEPARstart{W}{ireless} localization \cite{Yassin_2016} can be seen as one of the main applications in \ac{B5G} and \ac{6G} systems, in light of the goals defined in IMT-2030 \cite{02:00074} demonstrating how users can be localized only by a radio signal, which in fact can be a communication signal used for \ac{JCAS} applications.
While there are many types of information that can be extracted from radio signals for the purpose of localization, including finger-prints \cite{VoCST2016}, \ac{RSSI} \cite{Nic:RSSI}, \ac{AoA} \cite{Al-SadoonTAP2020}, or delay-based estimates of radio range \cite{ZengTSP2022}, with the large demand of high accuracy localization techniques, one has to consider the combination of different types of information to improve the performance.

Thus, \cite{Abreu_2007,Abreu_2008} proposed the \ac{SMDS} algorithm as an extension of the conventional \ac{MDS} method \cite{Torgerson_1952,Cox_2000}, which localizes target nodes by a combination of distance and angle measurements.

A consequence of this development and the large amount of new application is an increasing interest in the \acf{RBL} problem \cite{WangTSP2020,führling2024enablingnextgenerationv2xperception,Chepuri_2013,Bras_2016}, whose objective is to determine not only the location of targets, but also their shape and orientation, by defining a collection of points that models the target rigid body, which can be of interest in a variety of applications, such as navigation \cite{eckenhoff_2019}, collision detection \cite{Bruk_2023}, or vehicle path prediction \cite{Huang_2022}.
To name a few examples of the radio-based \ac{RBL} approach, which is the main subject of this article, is the method in \cite{führling2025robustegoisticrigidbody}, where a \ac{MDS} based approach was proposed that estimates the rigid body parameters efficiently in scenarios with incomplete observations, unaware of the targets shape.
Another example is \cite{Chen_2015}, in which a two-stage approach was used to estimate rotation, translation, angular velocity and translational velocity by range and Doppler measurements, making use of various \ac{WLS} minimization methods.
Finally, in \cite{führling20246drigidbodylocalization}, the problem was solved via \ac{GaBP}, estimating not only the stationary parameters, but also the velocity of the target. 

In view of the above, we propose in this article a \ac{SMDS}-based approach for \ac{RBL}, in which a rigid body can be estimated by a combination of distance and angle measurements that yields, the position, as well as the shape and orientation of the target.

The structure of the remainder of article is as follows.
First, a description of the rigid body system model and the \ac{SMDS} edge kernel is offered in Section \ref{sec:prior}.
Then, in Section \ref{sec:prop}, the proposed method for the estimation of the target rigid bodies' landmark points, translation, and orientation is introduced.
Finally, a comparison of the proposed scheme with the \ac{MDS}-based \ac{RBL} method, as well as an \ac{SMDS} approach that only utilizes distance measurements, is presented in Section \ref{sec:res}.

\vspace{-1ex}
\section{Rigid Body Localization System Model}
\label{sec:prior}
\vspace{-1ex}

\subsection{Rigid Body System Model}

\begin{figure}[H]
\centering
\includegraphics[width=\columnwidth]{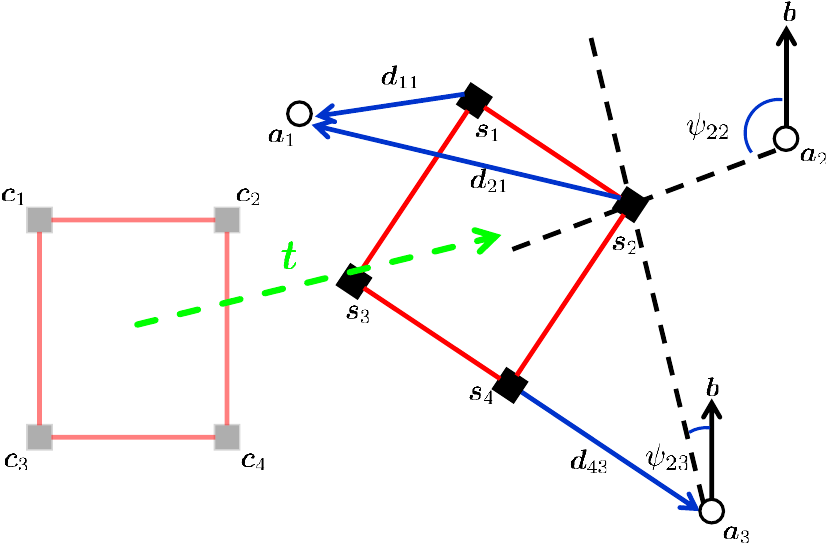}
\vspace{-4ex}
\caption{Illustration of an example \ac{RBL} scenario. The shape of the rigid body is defined by the distinct conformation matrices $\boldsymbol{C}$. The translation vector $\boldsymbol{t}$ between the two locations is depicted in green. All nodes are able to perform distance and angle of arrival measurements to each other, where angles are measured w.r.t. the vector $\bm{b}$.}
\label{fig:RB_sys}
\vspace{-2ex}
\end{figure}

Consider a collection $N$ of target points $\boldsymbol{c}_n\in\mathbb{R}^{2\times 1}$ in the \ac{2D} space\footnote{Due to the use of complex numbers, the approach is limited to 2D networks. The extension to 3D is trivial, but laborious and therefore omitted.}, with $n=\{1,\cdots,N\}$ that represent a rigid body, as shown in Figure \ref{fig:RB_sys}.
The structure and shape of said rigid body is therefore described by the corresponding conformation matrix $\boldsymbol{C}$ constructed by the column-wise collection of the vectors $\boldsymbol{c}_n$.
Then, consider the location $\bm{S}$ that the rigid body moved to from its original conformation, which can be modeled by the following relationship, as
\vspace{-0.5ex}
\begin{equation}
\label{eq:basic_model_one_body}
\boldsymbol{S}=\boldsymbol{Q}\cdot\boldsymbol{C}+\boldsymbol{t}\cdot\boldsymbol{1}_{N}^{\intercal},
\vspace{-0.5ex}
\end{equation}
where $\boldsymbol{t}\in\mathbb{R}^{2\times 1}$ is a translation vector given by the difference of the geometric centers of the body at the two locations, $\boldsymbol{1}_{N}$ is a column vector with $N$ entries all equal to $1$, and $\boldsymbol{Q}\in \mathbb{R}^{2\times 2}$ is a rotation matrix\footnote{Note that the rotation matrix is in the $SO(2)$ group, $i.e.$, $\bm{Q}^\intercal\bm{Q}=\bm{I}$ and $\text{det}(\bm{Q})=1$, which can easily be extended to 3D, by the yaw, pitch and roll angles.} determined by corresponding angle $\alpha$, namely
\vspace{-0.5ex}
\begin{equation}
\bm{Q} \triangleq \!\!\!
\begin{bmatrix}
\cos\alpha & -\sin\alpha \\
\sin\alpha & \cos\alpha
\end{bmatrix}\!\!.
\end{equation}

% \begin{eqnarray}
% \bm{Q} \triangleq \bm{Q}_{z}(\gamma)\,\bm{Q}_{y}(\beta)\,\bm{Q}_{x}(\alpha)&&\\
% %
% &&\hspace{-28ex}
% =\!\!\!\left[
% \begin{array}{@{}c@{\;\,}c@{\;\,}c@{}}
% \cos\gamma&-\sin\gamma& 0\\
% \sin\gamma& \cos\gamma& 0\\
% 0 	    & 0           & 1\\
% \end{array}\right]\!\!\!\cdot\!\!\!
% \left[
% \begin{array}{@{}c@{\;\,}c@{\;\,}c@{}}
% \cos\beta & 0           & \sin\beta\\
% 0			& 1			  & 0\\
% -\sin\beta& 0 		  & \cos\beta\\
% \end{array}\right]\!\!\!\cdot\!\!\!
% \left[
% \begin{array}{@{\,}c@{\;\,}c@{\;\,}c@{\!}}
% 1 			& 0			  & 0\\
% 0			& \cos\alpha& -\sin\alpha\\
% 0			& \sin\alpha& \cos\alpha\\
% \end{array}\right]\!\!.\nonumber
% \end{eqnarray}

In addition to the rigid body, there are $M$ anchor nodes $\bm{A}=[\bm{a}_1,\cdots,\bm{a}_m,\cdots,\bm{a}_M]\in\mathbb{R}^{2\times M}$ in the environment, where all nodes are capable of measuring distance and angle information to each other.

\subsection{SMDS Edge Kernel Construction}

In the network, a total of $T=N+M$ nodes are placed in \ac{2D} space, where the anchor nodes locations are known, while the location of the rigid bodies landmark points are unknown.
Following \cite{Abreu_2018}, the coordinates of the $i$-th node can be expressed as a complex number $x_{i} = a_{{x}_i}+jb_{x_i}$ such that the coordinates of all nodes can be collected in
\vspace{-0.5ex}
\begin{equation}
\label{EQ_ComplexCoordinateVector}
\boldsymbol{X} \triangleq
\left[
\begin{array}{c;{1pt/1pt}c}
\!\!\bm{A}\! & \!\bm{S}\!\!
\end{array}
\right]^\intercal
\in \mathbb{C}^{T\times 2}.
\vspace{-1ex}
\end{equation}

Next, consider the set of unique ascending index pairs $\mathcal{P} \triangleq \{(1,2),\cdots,(1,T),(2,3),\cdots,(T-1,T)\}$, such that each element $p \in \mathcal{P}$ corresponding to a pair of indices $(i,j)$ relates to the \emph{complex edge} $v_{p}$, defined as
\vspace{-0.5ex}
\begin{align}
\label{EQ_ComplexEdge}
v_{p}\!&= (a_{x_{j}}-a_{x_i}) + j (b_{x_{j}}-b_{x_{i}})  \nonumber \\
&= a_{v_{p}} + j b_{v_{p}} = d_{p} (\cos\theta_{p} + j\sin\theta_{p}),
\vspace{-0.5ex}
\end{align}
where $d_{p}$ is the Euclidean distance between the pair of nodes which is given by
\vspace{-0.5ex}
\begin{equation}
\label{Euclidean_Distance}
d_{p} \triangleq \Vert v_{p}\Vert = 
\Vert \bm{x}_i-\bm{x}_j \Vert,
\vspace{-0.5ex}
\end{equation}
where $\bm{x}_i$ denotes the $i-$th row of the node matrix $\bm{X}$.

In total there are  $P=\big({T\atop 2}\big)=T(T-1)/2$ complex edges, which, through the ascending index pairs, can be grouped as
\vspace{-1ex}
\begin{equation}
\vspace{-0.5ex}
\label{eq:edge_vec}
\boldsymbol{v}=\left[
\begin{array}{c;{1pt/1pt}c;{1pt/1pt}c}\!\!
\boldsymbol{v}_{_{\!\text{AA}}}^\intercal&
\boldsymbol{v}_{_{\!\text{AT}}}^\intercal&
\boldsymbol{v}_{_{\text{TT}}}^\intercal\!\!
\end{array}
\right]^\intercal = \mathbf{C}\cdot\boldsymbol{x}\in \mathbb{C}^{M},
\vspace{-0.5ex}
\end{equation}
where $\boldsymbol{v}_{_{\!\text{AA}}}$ represents the anchor-to-anchor, $\boldsymbol{v}_{_{\!\text{AT}}}$ the anchor-to-target and $\boldsymbol{v}_{_{\text{TT}}}$ the target-to-target complex edges and the coefficient matrix is given by
\begin{equation}
\label{EQ_C_MRC}
\mathbf{C} \triangleq
\left[
\begin{array}{c;{1pt/1pt}c;{1pt/1pt}c}\!\!
\mathbf{C}_{_{\!\text{AA}}}^\intercal &
\mathbf{C}_{_{\!\text{AT}}}^\intercal &
\mathbf{C}_{_{\!\text{TT}}}^\intercal \!\!
\end{array}
\right]^\intercal,
\vspace{-2ex}
\end{equation}
with
\begin{subequations}
\begin{equation}
\label{EQ_RankDefficientCoefficientMatrix_AA}
\mathbf{C}_{_{\!\text{AA}}} \!\triangleq\!
\left[
\begin{array}{@{}@{}c@{};{1pt/1pt}c@{};{1pt/1pt}c;{1pt/1pt}@{}c@{};{1pt/1pt}@{}c@{}@{}c}
\boldsymbol{1}_{_{M\!-1 \times 1}} & \multicolumn{4}{c}{-\mathbf{I}_{_{M\!-1 \times M\!-1}}} &
\multicolumn{1}{;{1pt/1pt}c}{\boldsymbol{0}_{_{M\!-1 \times N}}} \\[1ex]
\hdashline[1pt/1pt]
\boldsymbol{0}_{_{M\!-2 \times 1}} & \boldsymbol{1}_{_{M\!-2 \times 1}} &
\multicolumn{3}{c}{-\mathbf{I}_{_{M\!-1 \times M\!-1}}} &
\multicolumn{1}{;{1pt/1pt}c}{\boldsymbol{0}_{_{M\!-2 \times N}}} \\
\hdashline[1pt/1pt]
\multicolumn{2}{c;{1pt/1pt}}{\ddots} & \ddots & \multicolumn{2}{c}{\ddots} &
\multicolumn{1}{;{1pt/1pt}c}{\vdots} \\
\hdashline[1pt/1pt]
\multicolumn{3}{c;{1pt/1pt}}{\boldsymbol{0}_{_{1 \times M\!-2}}} &
\;1\; & -1 &
\multicolumn{1}{;{1pt/1pt}c}{\boldsymbol{0}_{_{1 \times N}}}
\end{array}
\right]\!\!,
\end{equation}

\begin{equation}
\label{EQ_RankDefficientCoefficientMatrix_AT}
\mathbf{C}_{_{\!\text{AT}}} \!\triangleq\!
\left[
\begin{array}{c;{1pt/1pt}c;{1pt/1pt}cccc}
\boldsymbol{1}_{_{N \times 1}} & \multicolumn{4}{c}{\boldsymbol{0}_{_{N \times M\!-1}}} &
\multicolumn{1}{;{1pt/1pt}c}{-\mathbf{I}_{_{N \times N}}} \\
\hdashline[1pt/1pt]
\boldsymbol{0}_{_{N \times 1}} & \boldsymbol{1}_{_{N \times 1}} &
\multicolumn{3}{c}{\boldsymbol{0}_{_{N \times M\!-2}}} &
\multicolumn{1}{;{1pt/1pt}c}{-\mathbf{I}_{_{N \times N}}} \\
\hdashline[1pt/1pt]
\ddots & \ddots & \multicolumn{1}{c}{\ddots} & \multicolumn{2}{;{1pt/1pt}c}{\ddots} &
\multicolumn{1}{;{1pt/1pt}c}{\vdots} \\
\hdashline[1pt/1pt]
\multicolumn{4}{c;{1pt/1pt}}{\boldsymbol{0}_{_{N \times M\!-1}}} &
\boldsymbol{1}_{_{N \times 1}} &
\multicolumn{1}{;{1pt/1pt}c}{-\mathbf{I}_{_{N \times N}}}
\end{array}
\right]\!\!,
\end{equation}

\begin{equation}
\label{EQ_RankDefficientCoefficientMatrix_TT}
\mathbf{C}_{_{\!\text{TT}}} \!\triangleq\!
\left[
\begin{array}{c@{};{1pt/1pt}c@{};{1pt/1pt}c@{};{1pt/1pt}c;{1pt/1pt}c;{1pt/1pt}c@{}}
\boldsymbol{0}_{_{N\!-1 \times M}} & \boldsymbol{1}_{_{N\!-1 \times 1}} &
\multicolumn{4}{c}{-\mathbf{I}_{_{N\!-1 \times N\!-1}}} \\
\hdashline[1pt/1pt]
\boldsymbol{0}_{_{N\!-2 \times M}} & \boldsymbol{0}_{_{N\!-2 \times 1}} &
\boldsymbol{1}_{_{N\!-2 \times 1}} & \multicolumn{3}{c}{-\mathbf{I}_{_{N\!-1 \times N\!-1}}} \\
\hdashline[1pt/1pt]
\vdots & \multicolumn{2}{c;{1pt/1pt}}{\ddots} & \ddots & \multicolumn{2}{c}{\ddots} \\
\hdashline[1pt/1pt]
\boldsymbol{0}_{_{1 \times M}} & \multicolumn{3}{c;{1pt/1pt}}{\boldsymbol{0}_{_{1 \times N\!-2}}} &
\;1\; & \!-1
\end{array}
\right]\!\!.
\end{equation}
\end{subequations}

To construct the kernel needed for the \ac{SMDS} reconstruction, the dissimilarity measure between two edges $v_m $ and $v_p$ can be defined by the complex product
\begin{align}
\label{EQ_KernelElement2}
\nonumber
\kappa_{mp} = v_m^* v_p &= d_{m} (\cos\theta_{m}\! -\! j\sin\theta_{m}) \cdot d_{p} (\cos\theta_{p}\! +\! j\sin\theta_{p}) \\
\nonumber
%&= d_{m} d_{p}(\cos\theta_{m}\cos\theta_{p}+\sin\theta_{m}\sin\theta_{p} \\
%\nonumber &\,\,\,\,\,\,\,\,-j(\cos\theta_{m}\sin\theta_{p}-\cos\theta_{p}\sin\theta_{m}))\\
%\nonumber
& = d_{m}d_{p}(\cos(\theta_{p}-\theta_{m}) - j\sin(\theta_{p}-\theta_{m}))
\\
& = d_{m}d_{p}(\cos\theta_{mp} - j\sin\theta_{mp}).
\end{align}

From the dissimilarities of all pairs, the complex kernel matrix can be constructed as
\vspace{-0.5ex}
{
 \setlength{\arraycolsep}{2pt}%
\begin{equation}
\label{EQ_ComplexKernelBlock}
\boldsymbol{\mathcal{K}}\! = \!\boldsymbol{v}^* \boldsymbol{v}^\intercal \!=\!\!
\left[
\begin{array}{c;{1pt/1pt}c;{1pt/1pt}c}
\!\boldsymbol{v}_{_{\!\text{AA}}}^* \boldsymbol{v}_{_{\!\text{AA}}}^\intercal\! &
\!\boldsymbol{v}_{_{\!\text{AA}}}^* \boldsymbol{v}_{_{\!\text{AT}}}^\intercal\! &
\!\boldsymbol{v}_{_{\!\text{AA}}}^* \boldsymbol{v}_{_{\!\text{TT}}}^\intercal\! \\
\hdashline[1pt/1pt]
\!\boldsymbol{v}_{_{\!\text{AT}}}^* \boldsymbol{v}_{_{\!\text{AA}}}^\intercal\! &
\!\boldsymbol{v}_{_{\!\text{AT}}}^* \boldsymbol{v}_{_{\!\text{AT}}}^\intercal\! &
\!\boldsymbol{v}_{_{\!\text{AT}}}^* \boldsymbol{v}_{_{\!\text{TT}}}^\intercal\! \\
\hdashline[1pt/1pt]
\!\boldsymbol{v}_{_{\!\text{TT}}}^* \boldsymbol{v}_{_{\!\text{AA}}}^\intercal\! &
\!\boldsymbol{v}_{_{\!\text{TT}}}^* \boldsymbol{v}_{_{\!\text{AT}}}^\intercal\! &
\!\boldsymbol{v}_{_{\!\text{TT}}}^* \boldsymbol{v}_{_{\!\text{TT}}}^\intercal\!
\end{array}
\right]\!\!
=\!\!
\left[
\begin{array}{c;{1pt/1pt}c;{1pt/1pt}c}
\!\boldsymbol{\mathcal{K}}_{\!\text{A}}\! &
\!\boldsymbol{\mathcal{K}}_1\! &
\!\boldsymbol{\mathcal{K}}_2\! \\
\hdashline[1pt/1pt]
\!\boldsymbol{\mathcal{K}}_1^\intercal\! &
\!\boldsymbol{\mathcal{K}}_3\! &
\!\boldsymbol{\mathcal{K}}_4\! \\
\hdashline[1pt/1pt]
\!\boldsymbol{\mathcal{K}}_2^\intercal\! &
\!\boldsymbol{\mathcal{K}}_4^\intercal\! &
\!\boldsymbol{\mathcal{K}}_{\!\text{T}}\!
\end{array}
\right]\!\!.
\end{equation}
}

As shown in the \ac{SotA}, the kernel $\boldsymbol{\mathcal{K}}$ is of rank one, such that the complex edge vector $\boldsymbol{v}$ can be estimated through a low rank truncation method, as used in the MDS and SMDS algorithms \cite{Torgerson_1952,Abreu_2007}, namely
\begin{equation}
\vspace{-0.5ex}
\label{EQ_LowRankComplexVectors}
\hat{\boldsymbol{v}} = \sqrt{\lambda}\, \boldsymbol{u},
\vspace{-0.5ex}
\end{equation}
where $(\lambda,\boldsymbol{u})$ is the largest eigenpair of ${\boldsymbol{\mathcal{K}}}$.

Finally, when the complete edge vector estimate $\hat{\boldsymbol{v}}$ is available, the corresponding coordinate vector estimate $\hat{\boldsymbol{x}}$ can be obtained by the inversion of \eqref{eq:edge_vec}, written as
\vspace{-0.5ex}
\begin{equation}
\label{EQ_V2XRankDeficient}
\hat{\boldsymbol{x}} = \C^{-1}\cdot \hat{\boldsymbol{v}}.
\vspace{-0.5ex}
\end{equation}
%

%%%%%%%%%%%%%%%%%%%%%%%%%%%%%%%%%%%%%
\section{Proposed method}
\label{sec:prop}

In light of the above, the proposed method consists of two steps. The first step offers a variation of the \ac{SMDS} framework, adapted to the rigid body scenario, which yields the estimated landmark point positions of the rigid body $\hat{\bm{S}}$, while the second step makes use of a standard least square minimization approach to estimate the translation ${\bm{t}}$ and rotation ${\bm{Q}}$ of the rigid body, given $\hat{\bm{S}}$.

\subsection{Rigid Body Estimation}
\label{sec:RB_est}

Inspired by the Turbo MRC \ac{SMDS} algorithm \cite{Abreu_2018}, which improves the classical \ac{SMDS} \cite{Abreu_2007} for independent target nodes by taking into account only a minor of the complex edge kernel, the same concept can be applied to the scenario of a rigid body, where not only the anchor-to-anchor measurements are known, but also the target-to-target measurements.
Thus, for the \ac{RBL} variation, a different minor of the complex edge kernel needs to be selected that takes into account the known information, which yields
\begin{equation}
\label{EQ_itCoopMRCSMDS1}
\begin{bmatrix}
\bm{\mathcal{K}}_1  \\
\bm{\mathcal{K}}_3  \\
\bm{\mathcal{K}}_4^\intercal
\end{bmatrix}
=
\begin{bmatrix}
\bm{v}^*_{AA} \\
\bm{v}^*_{AT} \\
\bm{v}^*_{TT}
\end{bmatrix}
\cdot
\begin{bmatrix}
\bm{v}^\intercal_{AT} 
\end{bmatrix},
\vspace{-0.5ex}
\end{equation}
which can be rearranged to solve for $\boldsymbol{v}_{_{\!\text{AT}}}$, as
\vspace{-0.5ex}
\begin{equation}
\label{EQ_TurboMRCSMDS_v}
\bm{\hat{v}}^{(n+1)}_{AT} 
=
\frac{\begin{bmatrix} \bm{\mathcal{K}}_1^\intercal & \bm{\mathcal{K}}_3^\intercal & \bm{\mathcal{K}}_4\end{bmatrix}}{\left\| \begin{bmatrix} \bm{{v}}_{AA} & \bm{\hat{v}}^{(n)}_{AT} & \bm{{v}}_{TT} \end{bmatrix}^\intercal \right\|^2}
\begin{bmatrix}
\bm{{v}}_{AA} \\
\bm{\hat{v}}^{(n)}_{AT} \\
\bm{{v}}_{TT}
\end{bmatrix}.
\vspace{-0.5ex}
\end{equation}

Moreover, an initial estimate $\boldsymbol{v}_{_{\!\text{AT}}}^{(0)}$ can easily be obtained by 
\vspace{-1ex}
\begin{equation}
\label{EQ_CooperativeMRCSMDS}
\bm{\hat{v}}_{AT}
=
\frac{\begin{bmatrix} \bm{\mathcal{K}}_1^\intercal & \bm{\mathcal{K}}_4 \end{bmatrix} }{\left\| \begin{bmatrix} \bm{{v}}_{AA}  & \bm{{v}}_{TT} \end{bmatrix}^\intercal \right\|^2}\begin{bmatrix}
\bm{{v}}_{AA}  \\
\bm{{v}}_{TT}
\end{bmatrix}.
\vspace{-1ex}
\end{equation}

The algorithm of equation \eqref{EQ_TurboMRCSMDS_v} exploits \emph{all} the extrinsic information contained in $\boldsymbol{\mathcal{K}}$, albeit by using only a \emph{smaller} portion of the latter.
Finally, as shown in \cite{Abreu_2018}, after estimating the vector $\bm{\hat{v}}_{AT}$, equation \eqref{EQ_V2XRankDeficient} can be used to reconstruct the rigid body landmark points $\hat{\bm{S}}$.

While the algorithm and the construction of the complex kernel matrix relies on the measurements of the distance and angle information, in some scenarios it might only be possible to collect the distance measurements.
In such circumstances, it is possible to estimate the angle information prior to the \ac{SMDS} algorithm such that first, a \ac{MDS} step is performed, followed by the reconstruction of the angles, which can then be used to construct the kernel.

\subsection{Rigid Body Parameter Estimation}
\label{sec:Param_Est}
With the estimate of the rigid bodies landmark points in hand, a least squares approach can be applied to estimate the rotation and translation parameters, as proposed in \cite{Chen_2015}.
The estimation of the rotation and translation from the relationship between two sets of points, $i.e.$, the transformed landmark point position $\hat{\mathbf{s}}_i$ and the original landmark point position $\mathbf{c}_i$, defined by the conformation matrix, is a well known problem in the \ac{SotA} \cite{Umeyama_1991,Eggert1997} and can be written as a minimization problem, given by

\vspace{-2ex}
\begin{equation}
    \begin{split}
& \argmin _{\mathbf{Q}, \mathbf{t}} \sum_{i=1}^N\left(\hat{\mathbf{s}}_i-\left(\mathbf{Q} \mathbf{c}_i+\mathbf{t}\right)\right)^\intercal \mathbf{W}_i\left(\hat{\mathbf{s}}_i-\left(\mathbf{Q} \mathbf{c}_i+\mathbf{t}\right)\right),\\
& \text { s.t. } \mathbf{Q} \in S O(K),
  \end{split}
\end{equation}
where $\mathbf{W}_i$ is the wheighing matrix, defined by the inverse of the covariance of $\hat{\mathbf{s}}_i$. 
To generalze the problem it can be rewritten by a non-negative scalar weighting, as
\begin{equation}
  \begin{split}
    & \argmin _{\mathbf{Q}, \mathbf{t}} \quad G=\sum_{i=1}^N w_i\left\|\hat{\mathbf{s}}_i-\left(\mathbf{Q} \mathbf{c}_i+\mathbf{t}\right)\right\|^2, \\
& \text { s.t. } \mathbf{Q} \in S O(K),
  \end{split}
\end{equation}
with the weighting matrix given by $\mathbf{W}_i=$ $w_i \mathbf{I}$.

Next, the weighted average values are given by
\begin{subequations}
\begin{equation}
  \overline{\mathbf{s}}=\sum_{i=1}^N w_i \hat{\mathbf{s}}_i / \sum_{i=1}^N w_i,
\end{equation}
\vspace{-1ex}
\begin{equation}
   \quad \overline{\mathbf{c}}=\sum_{i=1}^N w_i \mathbf{c}_i / \sum_{i=1}^N w_i .
\end{equation}
\end{subequations}

To find a closed form solution, the first step is to set the derivative of $G$ with respect to $\mathbf{t}$ to zero, which yields
\begin{equation}
  \label{eq:t_est}
  \mathbf{t}=\overline{\mathbf{s}}-\mathbf{Q} \overline{\mathbf{c}},
\end{equation}

Plugging \eqref{eq:t_est} back into the objective function $G$ and substituting $\tilde{\mathbf{s}}_i=\hat{\mathbf{s}}_i-\overline{\mathbf{s}}$ and $\tilde{\mathbf{c}}_i=\mathbf{c}_i-\overline{\mathbf{c}}$, $G$ can be rewritten as
\begin{equation}
  \begin{split}
    G & =\Sigma_{i=1}^N w_i\left\|\tilde{\mathbf{s}}_i-\mathbf{Q} \tilde{\mathbf{c}}_i\right\|^2 \\
& =-2 \Sigma_{i=1}^N w_i \tilde{\mathbf{s}}_i^\intercal \mathbf{Q} \tilde{\mathbf{c}}_i+\mathcal{P} \\
& =-2 \operatorname{trace}\left(\mathbf{Q} \Sigma_{i=1}^N w_i \tilde{\mathbf{c}}_i \tilde{\mathbf{s}}_i^\intercal\right)+\mathcal{P},
  \end{split}
\end{equation}
where $\mathcal{P}=\sum_{i=1}^N w_i\left(\left\|\tilde{\mathbf{s}}_i\right\|^2+\left\|\tilde{\mathbf{c}}_i\right\|^2\right)$ is an independent constant.
Since minimizing $G$ is equivalent to maximizing the trace of $\left(\mathbf{Q} \sum_{i=1}^N w_i \overline{\mathbf{c}}_i \overline{\mathbf{s}}_i^\intercal\right)$, the optimal solution \cite{Eggert1997} is for the rotation matrix is found by 
\begin{equation}
  \label{eq:Q_est}
  \mathbf{Q}=\mathbf{V} \operatorname{diag}\left(\left[\mathbf{1}_{K-1}^\intercal, \operatorname{det}\left(\mathbf{V} \mathbf{U}^\intercal\right)\right]^\intercal\right) \mathbf{U}^\intercal
\end{equation}
where the \ac{SVD} of $\sum_{i=1}^N w_i \overline{\mathbf{c}}_i \tilde{\mathbf{s}}_i^\intercal$ defines $\mathbf{U} \Sigma \mathbf{V}^\intercal$.
Finally, plugging \eqref{eq:Q_est} back into \eqref{eq:t_est} gives the solution for the translation vector.
%%%%%%%%%%%%%%%%%%%%%%%%%%%%%%%%%%%%%%%%%%%%%%%%%%%%5

\section{Performance Evaluation}
\label{sec:res}

In this section we provide numerical results that illustrate the performance of the proposed \ac{SMDS}-based \ac{RBL} method.
The region of interest consists of a $10$m-by-$10$m room equipped with $8$ anchor nodes, and a rigid body with $8$ landmark points, similar to the simplified version illustrated in Figure \ref{fig:RB_sys}.

The corresponding distance measurements are modeled as gamma-distributed random variables \cite{Papoulis_2002} with the mean given by the true distance and a standard deviation $\sigma$.
In turn, the \ac{AoA} measurement errors are Tikhonov-distributed \cite{Abreu_2008}, with concentration parameter $\rho \geq 0$ inversely proportional to the angular error variance.
Due to the non-linear relationship between $\rho$ and angular error variances, the influence of angular errors is captured by the quantity $\zeta_{\theta}$, defined as the bounding angle of the 90$^\text{th}$ centered percentile, $i.e.$
\vspace{-0.5ex}
\begin{equation}
\label{AngularError}
\zeta_\theta = \theta_\textup{B} \, \bigg| 
	\displaystyle\int_{-\theta_\textup{B}}^{\theta_\textup{B}}\!\!\!
		p_{\!_\Theta}(t;\rho)\,d t = 0.9,
\vspace{-0.5ex}
\end{equation}
where $p_{\!_\Theta}(t;\rho)$ denotes the central Tikhonov distribution \cite{Abreu_2008}.

The estimation errors are denoted by $\varepsilon$ and measured by the \ac{MSE} of the difference between the estimates and true values of the target parameters, $i.e.$
\begin{equation}
\varepsilon = \frac{1}{K}\sum_{k=1}^{K}|\hat{\boldsymbol{t}}^{(k)}-{\boldsymbol{t}}|_2^2,
\end{equation}
for the translation vector, where $\hat{\boldsymbol{t}}^{(k)}$ denotes the estimate at a $k$-th realization, which can also be applied to the estimates of the rotation matrix, calculating the results by averaging $K=10^3$ Monte-Carlo realizations. 

\begin{figure}[H]
\centering
\includegraphics[width=\columnwidth]{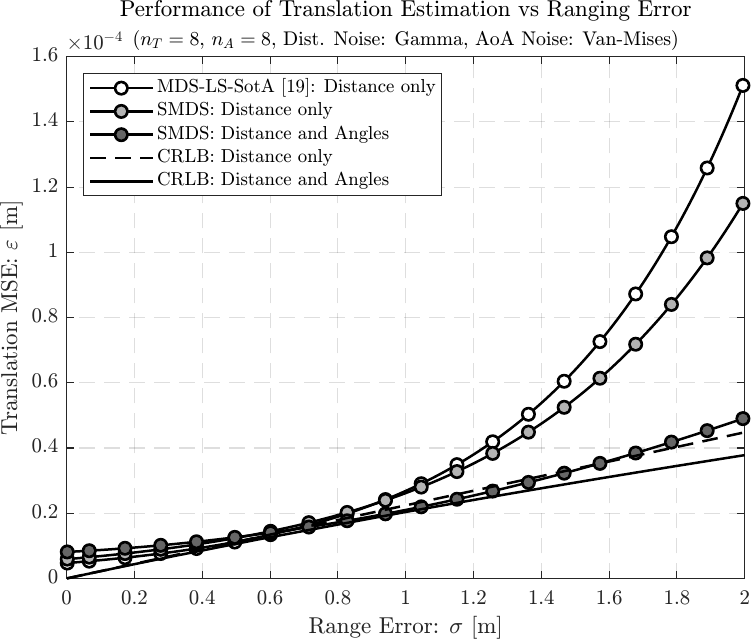}
\vspace{-4ex}
\caption{\ac{RMSE} of the translation estimate of the proposed method and the \ac{SotA}, over the range error $\sigma$.}
\label{fig:Tra}
\end{figure}
\vspace{-4ex}
\begin{figure}[H]
\centering
\includegraphics[width=\columnwidth]{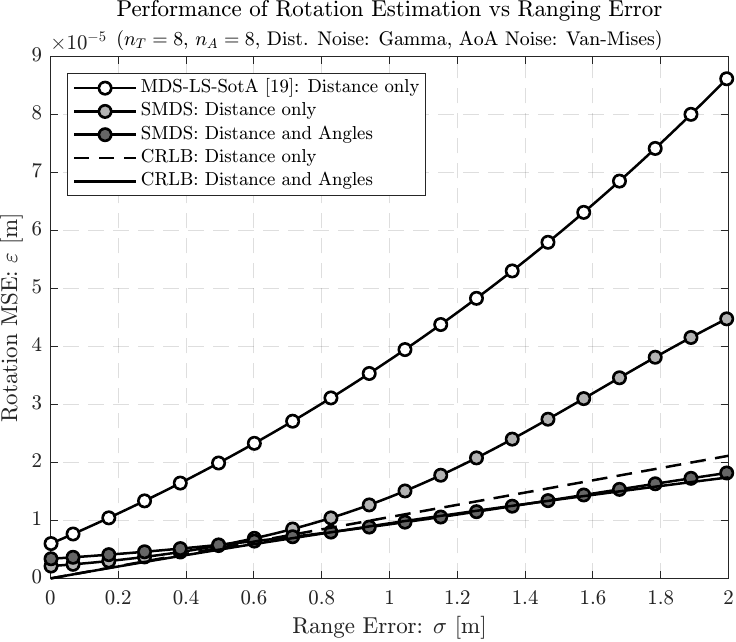}
\vspace{-4ex}
\caption{\ac{MSE} of the rotation estimate of the proposed method and the \ac{SotA}, over the range error $\sigma$.}
\label{fig:Rot}
\end{figure}

\subsection{Fundamental Limits}

To offer a detailed evaluation, the results of the proposed method are compared to the \acf{CRLB}, which offers a lower bound for the estimates.
To that extend, \cite{Nic_CRLB} proposed a generalized framework for fundamental limits in rigid body localization, where the \ac{FIM} can be calculated in an information-centric approach knowing the pairs of measurements, their corresponding type, $i.e.$, distance or \ac{AoA} measurements and the type of error the measurement is subject to.

Following \cite{Nic_CRLB}, the \ac{FIM} for the translation vector and the rotation matrix can be construced as

\begin{equation}
\label{eq:FIM_t_final}
\begin{split}
\mathbf{F}_{\bm{t}}&=\!\!\!\sum_{(n, a) \in \mathcal{P}_d}\lambda_{na}g' \big|_{\mathbf{t}}^{d}\Big(g' \big|_{\mathbf{t}}^{d}\Big)^\intercal+\!\!\sum_{(n, a) \in \mathcal{P}_\psi}\lambda_{na}g' \big|_{\mathbf{t}}^{\psi}\Big(g' \big|_{\mathbf{t}}^{\psi}\Big)^\intercal,
\end{split}
\end{equation}

\begin{equation}
\begin{split}   
\mathbf{F}_{\bm{Q}}&=\!\!\!\sum_{(n, a) \in \mathcal{P}_d}\lambda_{na}g' \big|_{\mathbf{Q}}^{d}\Big(g' \big|_{\mathbf{Q}}^{d}\Big)^\intercal+\!\!\sum_{(n, a) \in \mathcal{P}_\psi}\lambda_{na}g' \big|_{\mathbf{Q}}^{\psi}\Big(g' \big|_{\mathbf{Q}}^{\psi}\Big)^\intercal,
\end{split}
\end{equation}
where $g'$ indicates the information gradient of the respective parameter and information type and $\lambda$ denotes the information intensity defined by the type of eror distribution, with the corresponding derivations for the specific parameters found in \cite[Appendix A \& B]{Nic_CRLB}.
 
\subsection{Numerical Results}
Figure \ref{fig:Tra} and \ref{fig:Rot} show the result of the translation and rotation estimation for an \ac{MDS}-based approach, an \ac{SMDS}-based approach, where the angles are obtained by using only the distance measurements, and a full \ac{SMDS} approach compared to the corresponding \acp{CRLB}.  
It can be observed that the distance only \ac{SMDS} approach is performing better than the \ac{MDS} approach, while not performing as good as the full \ac{SMDS} approach, which is also expected, since the full \ac{SMDS} approach uses both distance and angle measurements.

In Figure \ref{fig:Tra} it can be observed that in general, the full \ac{SMDS} yields the best results with estimates close to the \ac{CRLB}, while the \ac{SMDS} approach with only distance measurements is performing better than the \ac{MDS} approach, and slightly better that the full \ac{SMDS} in small range error regimes, which is expected, since in \cite{Abreu_2007} it was shown that the \ac{SMDS} approach is not optimal in the small range error regime.

In Figure \ref{fig:Rot} similar properties can be observed, where the full \ac{SMDS} approach is performing best, while the \ac{SMDS} approach with only distance measurements is performing much better than the \ac{MDS} approach, with the full \ac{SMDS} approach performing close to the \ac{CRLB} over the whole error range.

%%%%%%%%%%%%%%%%%%%%%%%%%%%%%%%%%%%%
\section{Conclusion}

We proposed a novel \ac{SMDS}-based \ac{RBL} algorithm, which enables a rigid bodies relative translation (effective distance) and orientation (relative rotation) to be detected by a set anchor nodes the  of another body, based only on a set of measurements of the distance and \ac{AoA} information between sensors of the target an dthe anchor landmark points.
A key point of the proposed method is that compared to conventional \ac{SMDS}, the solution can be found in an iterative manner, while only a minor part of the complex edge kernel is considered that depends on the known noise-free measurements, which in the proposed scenario are anchor-to-anchor and target-to-target distances and angles.
Simulation results illustrate the good performance of the proposed technique in terms of \ac{MSE} as a function of the measurement error, reaching the fundamental limit illustrated by the \ac{CRLB}.
%
%Additionally, it was shown that when only distance measurements are available, it is possible to first perform classical \ac{MDS} to find the corresponding angles between the nodes, followed by the \ac{SMDS} approach, which clearly outperforms the \ac{MDS} only parameter estimation.

%-----------------REFERENCES---------------------------

\newpage

\bibliographystyle{IEEEtran}
\bibliography{IEEEabrv,ref_paper.bib}

\vspace{-2ex}

\end{document}